\documentclass[]{aa}

\usepackage{graphicx}
\usepackage{txfonts}
\usepackage[colorlinks=true,citecolor=blue,linkcolor=black]{hyperref}
\usepackage{siunitx}
\usepackage{xcolor}
\usepackage{float}
\usepackage{bm}
\usepackage{soul,xcolor}
\setstcolor{red}
\definecolor{alizarin}{rgb}{0.82, 0.1, 0.26}

\newcommand{\oiii}{\textup{[O\,\textsc{iii}]}}
\newcommand{\nii}{\textup{[N\,\textsc{ii}]}}
\newcommand{\sii}{\textup{[S\,\textsc{ii}]}}
\newcommand{\oi}{\textup{[O\,\textsc{i}]}}
\newcommand{\siii}{\textup{[S\,\textsc{iii}]}}

\newcommand{\hii}{\textup{H}\,\textsc{ii}}
\newcommand{\ha}{\textup{H}\ensuremath{\alpha}}
\newcommand{\hb}{\textup{H}\ensuremath{\beta}}

\defcitealias{Congiu2023}{C23}

\begin{document} 
\title{Machine learning the gap between real and simulated nebulae}

\subtitle{A domain-adaptation approach to classify ionised nebulae in nearby galaxies}
\titlerunning{Classifying ionised clouds with domain adaptation}

\author{Francesco~Belfiore\inst{\ref{arcetri}}
\and
Michele Ginolfi\inst{\ref{arcetri}, \ref{UNIFI}}
\and
Guillermo Blanc\inst{\ref{Carnegie}, \ref{unichile}}
\and
Mederic Boquien\inst{\ref{nice}}
\and 
Melanie Chevance\inst{\ref{ita}}
\and
Enrico Congiu\inst{\ref{eso_chile}}
\and
Simon C.\ O.\ Glover\inst{\ref{ita}}
\and
Brent Groves\inst{\ref{icrar}}
\and
Ralf S.\ Klessen \inst{\ref{ita},\ref{iwr},\ref{cfa},\ref{rad}}
\and
J. Eduardo M\'endez-Delgado\inst{\ref{ari}}
\and
Thomas~G.~Williams \inst{\ref{oxford}}
}

\institute{INAF -- Arcetri Astrophysical Observatory, Largo E. Fermi 5, I-50125, Florence, Italy\label{arcetri}. \email{francesco.belfiore@inaf.it}
\and Universit\`a di Firenze, Dipartimento di Fisica e Astronomia, via G. Sansone 1, 50019 Sesto Fiorentino, Florence, Italy\label{UNIFI}
\and The Observatories of the Carnegie Institution for Science, 813 Santa Barbara St, Pasadena, CA 91101, USA\label{Carnegie}
\and Departamento de Astronomía, Universidad de Chile, Camino del Observatorio 1515, Las Condes, Santiago, Chile\label{unichile}
\and  Université Côte d’Azur, Observatoire de la Côte d’Azur, CNRS, Laboratoire Lagrange, 06000, Nice, France\label{nice}
\and Universit\"at Heidelberg, Zentrum f\"ur Astronomie, Institut f\"ur Theoretische Astrophysik, Albert-Ueberle-Stra{\ss}e 2, D-69120, Heidelberg, Germany\label{ita}
\and European Southern Observatory (ESO), Alonso de C\'ordova 3107, Casilla 19, Santiago 19001, Chile \label{eso_chile}
\and International Centre for Radio Astronomy Research, University of Western Australia, 7 Fairway, Crawley, 6009 WA, Australia \label{icrar}
\and Astronomisches Rechen-Institut, Zentrum f\"ur Astronomie der Universit\"at Heidelberg, M\"onchhofstra{\ss}e 12-14, D-69120 Heidelberg, Germany\label{ari}
\and Universit\"{a}t Heidelberg, Interdisziplin\"{a}res Zentrum f\"{u}r Wissenschaftliches Rechnen, Im Neuenheimer Feld 225, 69120 Heidelberg, Germany \label{iwr}
\and Center for Astrophysics $|$ Harvard \& Smithsonian, 60 Garden Street, Cambridge, MA 02138, U.S.A. \label{cfa}
\and Radcliffe Institute for Advanced Studies at Harvard University, 10 Garden Street, Cambridge, MA 02138, U.S.A. \label{rad}
\and Sub-department of Astrophysics, Department of Physics, University of Oxford, Keble Road, Oxford OX1 3RH, UK\label{oxford}
}

\date{Received XXX; accepted XXX}

 
\abstract{
Classifying ionised nebulae in nearby galaxies is crucial to studying stellar feedback mechanisms and understanding the physical conditions of the interstellar medium. 
This classification task is generally performed by comparing observed line ratios with photoionisation simulations of different types of nebulae (\hii\ regions, planetary nebulae, and supernova remnants). However, due to simplifying assumptions, such simulations are generally unable to fully reproduce the line ratios in observed nebulae. This discrepancy limits the performance of the classical machine-learning approach, where a model is trained on the simulated data and then used to classify real nebulae. For this study, we used a domain-adversarial neural network (DANN) to bridge the gap between photoionisation models (source domain) and observed ionised nebulae from the PHANGS-MUSE survey (target domain). The DANN is an example of a domain-adaptation algorithm, whose goal is to maximise the performance of a model trained on labelled data in the source domain on an unlabelled target domain by extracting domain-invariant features.
Our results indicate a significant improvement in classification performance in the target domain when employing the DANN framework compared to a classical neural network (NN) classifier. 
Additionally, we investigated the impact of adding noise to the source dataset, finding that noise injection acts as a form of regularisation, further enhancing the performances of both the NN and DANN models on the observational data. The combined use of domain adaptation and noise injection improved the classification accuracy in the target domain by 23\%. This study highlights the potential of domain adaptation methods in tackling the domain-shift challenge when using theoretical models to train machine-learning pipelines in astronomy.
}

\authorrunning{F. Belfiore et al.}
\maketitle


\section{Introduction} 
\label{sec:intro}

Ionised nebulae are cosmic spotlights that provide us with a view of the properties of nearby as well as distant galaxies. \hii\ regions, where young stars ionise the surrounding interstellar medium (ISM), are key probes of stellar feedback \citep{Lopez2014, Pellegrini2020, Barnes2020, Chevance2023} and tracers of the physical conditions and chemical abundances of the ISM \citep{Peimbert2017, Kewley2019, Maiolino2019}. Given their importance for a wider range of astrophysical studies, a large body of work has been dedicated to the classification of ionised ISM emission in galaxies in order to separate \hii\ regions from other ionising sources, including active galactic nuclei \citep{Kewley2001, Kauffmann2003a}, shocks \citep{Allen2008}, hot evolved stars \citep{Binette2012, Byler2019}, and diffuse ionised gas \citep{CidFernandes2011, Zhang2017, Belfiore2022}. 

Observations capable of resolving individual nebulae are particularly useful to test the physics of the ionised ISM and refine our classification strategies. The required resolution ($\sim$100~pc) can be obtained using ground-based observations in galaxies closer than $\sim$ 20 Mpc. Several surveys (PHANGS, \citealt{Emsellem2022}; SIGNALS, \citealt{Rousseau-Nepton2019}; and MAUVE, e.g. \citealt{Watts2024}) have used wide-field integral field spectrography to map ionised gas in nearby galaxies at such a high resolution to study the physics of the small-scale star formation cycle in its galactic context. In these datasets, one can confidently separate high-density nebular structures from the ionised gas background \citep{Roth2018, McLeod2021, Santoro2022, Belfiore2022}. The classification task for ionised ISM structures is therefore simplified to distinguishing between \hii\ regions, planetary nebulae (PNe), and supernova remnants (SNRs). Recent literature has leveraged both Bayesian methods and machine-learning algorithms for this classification problem \citep{Kopsacheili2020, Micheva2022, Congiu2023, Rhea2023}, and to infer the physical properties of \hii\ regions \citep{Kang2022, Kang2023}.


This classical approach to classifying ionised nebulae is based on comparing measured emission-line ratios with those predicted by photoionisation simulations\footnote{To avoid confusion, in this work we refer to photoionisation models as `simulations', and use the term `model' only to refer to a machine-learning model.} of the different classes of objects. Widely used line-ratio diagnostic diagrams, such as the Baldwin-Phillips-Terlevich (BPT, \citealt{Baldwin1981}) diagram, represent useful simplified projections. However, in the BPT two-dimensional planes, substantial overlap exists between the locations inhabited by the different classes, leading to ambiguous classifications. Such limitations can be overcome by using Bayesian methods, where the full set of available line ratios is used to infer the final classification  \citep[][hereafter C23]{Congiu2023}. 

Nonetheless, the classification approach based on comparison with simulations is fundamentally limited because the current generation of simulations cannot fully match the data across all observed line ratios. These limitations may be due to the simple geometries assumed in most simulations. In the case of  \hii\ regions, for example, density (or temperature) substructures have been invoked to explain the measured discrepancy between abundances measured from recombination lines and collisionally excited ones \citep{Mendez-Delgado2023, Mendez-Delgado2023a}. Limitations in geometry also do not take into account the possible presence of low-density paths, which may lead to the escape of ionising photons. Real nebulae may also feature multiple ionisation sources with non-trivial spatial distributions. While more complex three-dimensional models of \hii\ regions have been constructed \citep{Bisbas2009, Walch2012, Jin2022a}, such approaches cannot yet produce a sufficiently large and varied dataset for training machine-learning models.

\begin{figure}[ht!]
\centering
\includegraphics[width=0.5\textwidth,trim=0 20 0 10, clip]{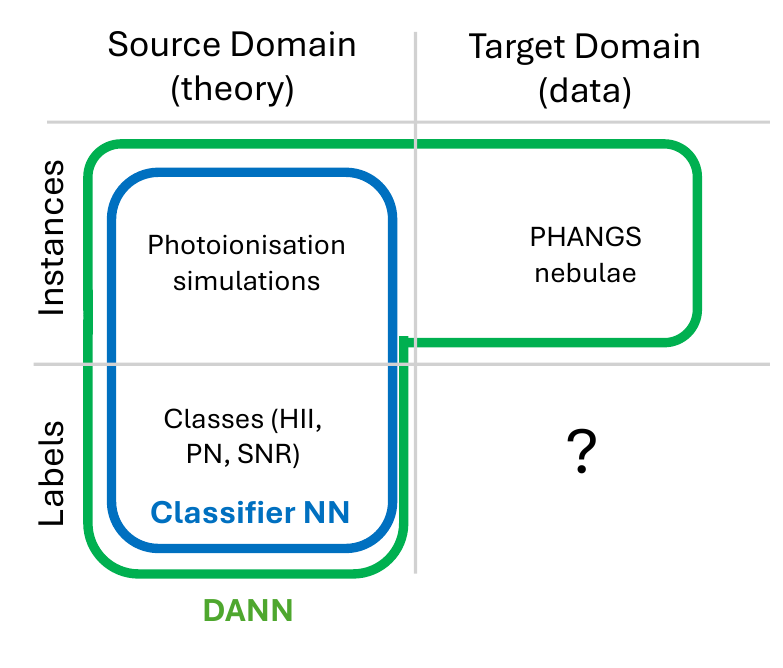}
\caption{Schematic representation of the domain-adaptation problem tackled in this work. The task consists in inferring the classification of PHANGS nebulae, given a set of photoionisation simulations from nebulae of different classes. The source domain (theory) contains the labels, while the target domain (data) is unlabelled. A classical, supervised machine-learning algorithm (e.g. a classifier neural network, in blue) is trained on the instances and labels of the source domain, but may perform poorly on the target domain, because of the differences between simulated and real data. In this work we employed a DANN, a domain-adaptation algorithm that uses as input (in green) both the labelled source domain and the unlabelled target domain instances. This approach allows the model to learn domain-invariant features and perform better on the target domain. }
\label{fig:explain_da}
\end{figure}

This mismatch between theoretical simulations and empirical data leads to what in the language of machine learning is referred to as a `domain adaptation problem'. Here, a model needs to be trained on a labelled set of data (instances) and then applied to a different domain of interest in which the instances are unlabelled. The two domains are referred to as the `source domain' (the one that includes labels, in our case the theory), and the `target domain' (the unlabelled, empirical data). A schematic representation of the situation is presented in Fig. \ref{fig:explain_da}. A classical, supervised machine-learning approach (e.g. a classifier neural network) relies on training a model on the source domain, the only one for which we have labels, and then applying the same trained model for inference on the unlabelled target domain. However, because of the differences between the two domains, this approach generally leads to models that do not perform well on the target domain.

In domain adaptation, on the other hand, an algorithm is trained on both the labelled source and unlabelled target domain instances. By learning so-called `domain-invariant' features, shared by both source and target domain, such an algorithm can perform significantly better, effectively reconciling the differences between theory and empirical data. In this work we will use a DANN (domain adversarial neural network, \citealt{Ganin2016}) to perform the required domain adaptation. 

Domain adaptation has found some early applications in astronomy, especially in the context of classifying images. The task of recognising galaxy mergers in large imaging surveys, for example, has attracted some attention in the literature. Since mergers are rare, training on simulations is mandatory to obtain a suitably large dataset. Early studies have demonstrated, however, that training on simulations without taking observational effects into account results in poor performance \citep{Bottrell2019, Pearson2019}. Using domain-adaptation algorithms, on the other hand, leads to substantial improvements \citep{Ciprijanovic2021}.

In this work, we test the ability of domain-adaptation techniques to handle the problem of classification of ionised nebulae based on their observed line ratios.  For this proof-of-concept analysis, we focus on the well-studied problem of classifying ionised nebulae in nearby galaxies into \hii\ regions, PNe, and SNRs using the catalogue of  \citetalias{Congiu2023}, based on optical integral field spectroscopy from the PHANGS-MUSE survey. 
We adopt standard machine-learning terminology, referring to the vector containing the line ratios for each nebula as an `instance', the individual line ratios as `features', and the classifications as `labels'. We present the implementation of a DANN and compare its performance against a classical artificial neural network (NN) classifier, in both the source and target domains. Additionally, we explore the impact of introducing noise to the source domain in order to more closely mimic realistic observations.

In detail, in Sec. \ref{sec:data} we present the datasets associated with the source and target domains. In Sec. \ref{sec:methods} we describe our NN and DANN model architectures.  In Sec. \ref{sec:res} we present the results obtained with both the classical NN and the DANN classifier, compare their performance, and discuss the impact of our proof-of-concept analysis. In Sec. \ref{sec:conclusions} we present a summary of our results.

\section{Data} 
\label{sec:data}

\subsection{Source domain: photoionisation simulations}
\label{sec:models}

The source domain in this work is represented by simulations of ionised nebulae generated by the photoionisation code CLOUDY \citep{Ferland1994, Ferland2017} and photoionisation and shock-modelling code MAPPINGS \citep{Sutherland2018}. These codes model the state of the ionised gas from first principles and predict the emerging spectrum of a nebula, but the resulting simulations suffer from limitations due to simplifying assumptions (all simulations are one-dimensional and homogeneous) and from missing or inaccurate physics (e.g. in their input ionising stellar spectra). Moreover, while simulation grids can be run for a wide range of input parameters (e.g. density, chemical abundances, ionisation parameters), real ionised nebulae generally only cover a part of this parameter space. In previous works, such unrealistic simulations were often excluded from the grid `by hand' \citep{Amayo2021} to reduce their impact on inference. Alternatively, more physical simulations can be constructed by coupling photoionisation models with physical models for the time evolution of ionised nebulae \citep{Pellegrini2020a, Pellegrini2020}. Considering these factors, we expect simulations to show some mismatch with respect to observations of ionised nebulae, motivating our domain-adaptation approach.

For ease of comparison with the classifications performed by \citetalias{Congiu2023} we selected similar grids of simulations for \hii\ regions, PNe, and shocks. Shock simulations were used to identify SNRs, although regions shocked by active galactic nuclei or shock networks in the diffuse ionised gas are also compatible with these simulations. Nonetheless, in this work we employ the term shock and SNR interchangeably. We considered a slightly more extended grid than \citetalias{Congiu2023} by including a wider range of possible values of N/O for \hii\ regions and a wider range of possible densities for PNe. These parameters directly affect the line ratios of interest in this work, and we preferred to avoid `by-hand' model pruning since the domain adaptation step should be able to account for any potential shifts between models and data.

The photoionisation simulations adopted in this work were taken from \cite{Perez-Montero2014} for \hii\ regions, \cite{Delgado-Inglada2014} for PNe, and \cite{Allen2008} for shocks. The range of simulation input parameters selected for our grids are summarised in Table \ref{tab:models}. The line fluxes for these simulations were retrieved from the Mexican Million Models database \citep{Morisset2015}.\footnote{\url{https://sites.google.com/site/mexicanmillionmodels}, for \hii\ regions and PNe and \url{http://3mdb.astro.unam.mx:3686/} for shocks.}
We considered the emission line fluxes of a set of key optical emission lines, namely \ha, \hb, \oiii $\lambda$5007, \nii$\lambda$6584, \sii $\lambda$6717, \sii $\lambda$6731, and \oi $\lambda$6300. The choice of emission lines to consider was motivated by their observability, as discussed in the next section. 

In our machine-learning analyses, we randomly selected a subset of simulations so that each class was equally represented, leading to a total of 2016 instances. We selected a balanced set of instances for the source domain even though the target domain is likely to be unbalanced, since observational samples generally contain a majority of \hii\ regions.

\begin{figure*}[ht!]
\centering
\includegraphics[width=0.98\textwidth,trim=0 0 0 10, clip]{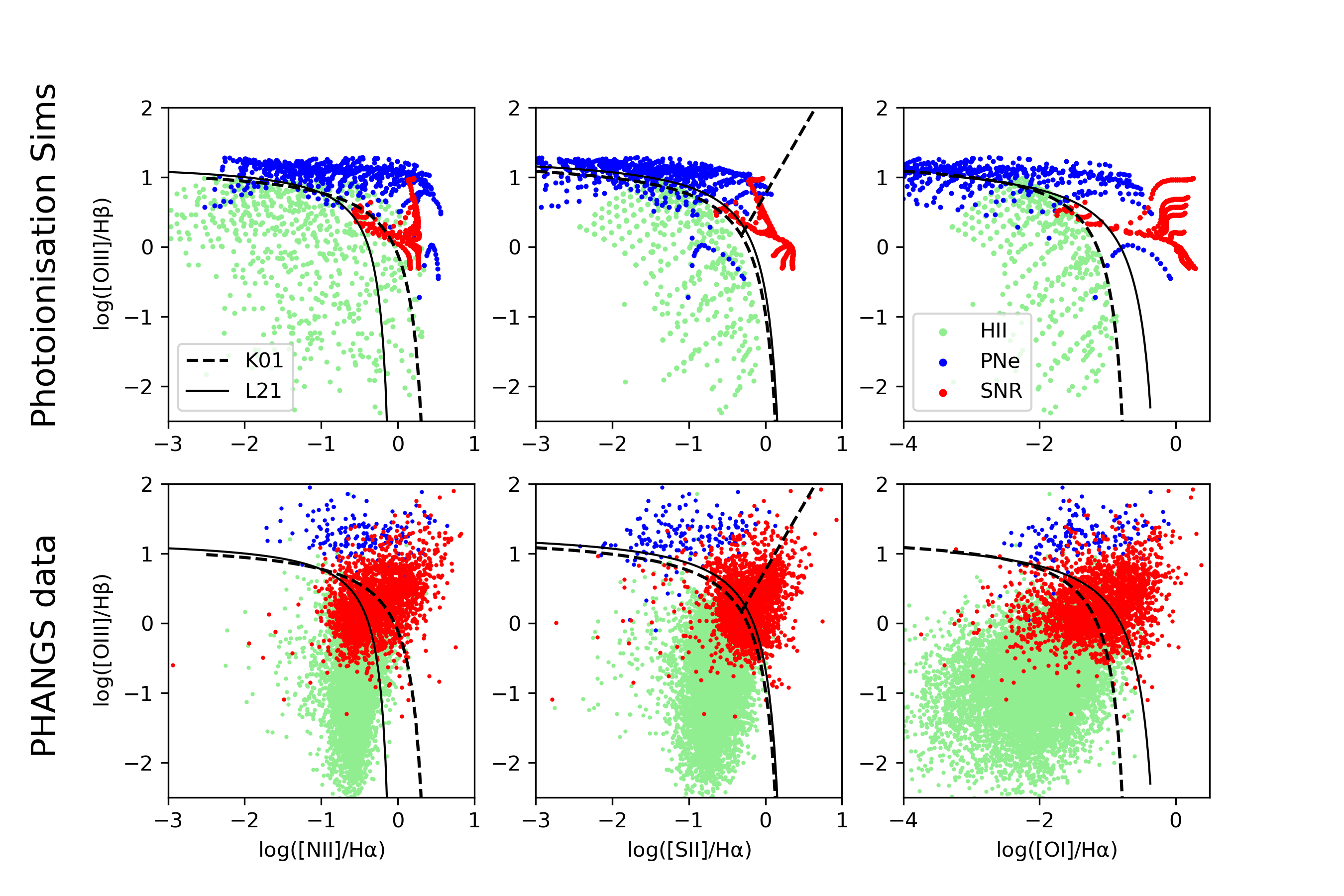}
\caption{Location in three BPT diagrams of photoionisation simulations (top row) and ionised nebulae from PHANGS-MUSE (\citetalias{Congiu2023}, bottom row). In each panel, the colour-coding represents class labels (green: \hii\ regions, blue: PNe, and red: SNRs). The class labels for the observed nebulae are based on the Bayesian methodology of \citetalias{Congiu2023} and are never used in the training, as we treat the target domain as unlabelled. Only the subset of the \citetalias{Congiu2023} catalogue used to train the classification algorithm presented in this work is shown. In each BPT diagram we show the demarcation lines of \cite{Law2021} (L21, solid line) and  \cite{Kewley2001, Kewley2006} (K01, dashed line). The \cite{Kauffmann2003a} line in the \nii-BPT is almost coincident with the \cite{Law2021} line. 
 The figure demonstrates that there is only a limited overlap in parameter space between simulations and data.}
\label{fig:data}
\end{figure*}

In Fig. \ref{fig:data} (top row) we show the position of the subsample of selected simulations in the classical BPT diagnostic diagrams, together with demarcation lines from the literature  \citep{Kewley2001, Kewley2006, Law2021}. The \cite{Kauffmann2003a} corresponds closely to the \cite{Law2021} in the \nii-BPT.
\hii\ regions occupy the bottom-left corner of these diagrams, as expected, and generally lie below the \cite{Law2021} line, except in the \nii-BPT, where \hii\ regions with non-canonical N/O ratios extend to the top right of the demarcation. PNe populate preferentially the top left of the BPT diagrams, lying mostly above the \hii\ demarcation lines except in the \oi-BPT. Shocks populate the region of the diagrams sometimes associated with low-ionisation emission-line regions (LIERs, \citealt{Belfiore2016}), and are better separated from other classes in the \sii\ and \oi-BPT.

\begin{table}[]
	\centering
 \caption{Summary of the photoionisation simulation parameters}
	\begin{tabular}{l c}
    \hline
    \multicolumn{2}{c}{\hii\ regions} \\
    \hline
    Reference  & \cite{Perez-Montero2014} \\
    Code        & Cloudy v13 \\
    12+log(O/H) & [7.1 ,9.1]\\
    ionisation parameter & [-4 ,-1.5]\\
    log(N/O) & [-2, 0]\\
    C tied to & O \\
	\hline
    \hline
	\multicolumn{2}{c}{Planetary Nebulae}\\
    \hline
    Reference & \cite{Delgado-Inglada2014} \\
    Code & Cloudy v17.01 \\
    log(Luminosity) & [200, 18000] \\
    Temperature & [5, 18]$\times 10^4$ K\\
    SED & Rauch \\
    Dust depletion & Yes \\
    log($n_e/\mathrm{cm^{-3}}$) & [1,5] \\
    Inner radius & $10^{17} ~ \mathrm{cm}$ \\
    Termination & Matter bounded M40 \\
    Abundances & solar \\
	\hline
    \hline
    \multicolumn{2}{c}{Shocks (Supernova Remnants)}\\
	\hline
    Reference & \cite{Allen2008} \\
    Code & MAPPINGS III \\
    Model set  & w/o precursor \\
    pre-schock density & 1 $\rm cm^{-3}$\\
    Magnetic param. & [0.0001, 10] $\rm \mu G ~cm^{3/2} $\\
    Shock velocity & [100, 1000] $\rm km~s^{-1}$\\
    Abundances & solar \\
    \hline
	\end{tabular}
	
	\label{tab:models}
\end{table}

\subsection{Target domain: PHANGS-MUSE ionised nebulae}
\label{sec:nebular}

We applied our classification task to the catalogue of ionised nebulae presented in \citetalias{Congiu2023} and retrieved from the Canadian Astronomical Society Data Centre\footnote{\url{https://www.canfar.net/storage/vault/list/phangs/RELEASES/Classified_Nebulae_Congiu2023}}. This catalogue contains emission-line fluxes for 40920 nebulae across a sample of 19 nearby ($D< 20$ Mpc) star-forming galaxies observed by the PHANGS-MUSE programme \citep{Emsellem2022}. Details on the catalogue generation are given in \citetalias{Congiu2023}, and we summarise here the key steps. First, a detection image was generated using a combination of line maps for the \oiii$\lambda$5007, \ha, and \sii $\lambda$6717,31 emission lines. The CLUMPFIND \citep{Williams1994} algorithm was run on this image to produce a first iteration of the segmentation maps. This segmentation was refined by identifying the isophotes containing 90\% of the region flux (after subtracting a local background), which was then chosen as the new region boundary. The catalogue was then cleaned to reject false detections by applying cuts in signal-to-noise ratio (S/N) and size. The line fluxes for each nebula were computed by fitting the integrated spectrum extracted from the datacube. Finally, a local background extracted in an annulus around the region was subtracted from the nebular flux to account for the contamination of diffuse ionised gas (DIG) along the line of sight. While this latter correction can be significantly uncertain for fainter nebulae, we chose to nonetheless use the background-subtracted fluxes for ease of comparison with the classifications performed by \citetalias{Congiu2023}.


\citetalias{Congiu2023} classified ionised gas nebulae using a Bayesian approach based on a data-model comparison across a set of the brightest emission lines in the MUSE wavelength range (\ha, \hb, \oiii $\lambda$5007, \sii $\lambda$6717, \sii $\lambda$6731, and \oi $\lambda$6300, and \siii$\lambda$ 9068). However, in this work we considered the nebulae from the \citetalias{Congiu2023} catalogue as an unlabelled target domain, that is, we did not use the class labels during training. We nonetheless used the labels inferred by \citetalias{Congiu2023} as a post-facto check on the performance of our classifications. 

From the \citetalias{Congiu2023} catalogue we selected a sub-sample of nebulae where the background-subtracted nebular line flux is larger than its error in the \ha, \hb, \nii, and \oiii\ emission lines. For the sake of simplicity we also removed nebulae that were classified as `ambiguous' by \citetalias{Congiu2023}, which were usually objects with a very low S/N. This cut reduced the sample to 17431 objects. Our sample is strongly unbalanced in terms of class membership, comprising of 14042 (80.5\%) \hii\ regions, 270 (1.5\%) PNe, and 3119 (18\%) SNRs.

In this work we did not experiment with probabilistic models \citep{Kang2023}, so our machine-learning approach is not able to account for uncertainties and upper limits on data. Unfortunately, this fact represents a significant limitation for the catalogue under study. In particular, $\sim50 \% $ of the objects classified as PNe by \citetalias{Congiu2023} are not detected in \sii\ or \oi, and only 23\% are detected in both. To avoid further curtailment of our sample of PNe, we decided to retain the objects with undetected \sii\ or \oi, and used the measured 1$\sigma$ error as their flux. We do not expect this choice to have a significant effect on our classification because, as evident from Fig. \ref{fig:data} (upper row), the PNe models cover a wide range of \sii/\ha\ and \oi/\ha. We also opted to exclude the \siii$\lambda$9068 line in our analysis because of its limited detection statistics and because the \siii$\lambda$9068 line has so far been difficult to reproduce in photoionisation models of \hii\ regions \citep{Mingozzi2019}. 

Finally, we corrected the observed data for dust attenuation along the line of sight because photoionisation models do not include any foreground dust screen model. We calculated the dust correction using the Balmer decrement, assuming case B recombination at a temperature of $T = 10^4 $ K, and the attenuation law of \cite{O'Donnell1994}. We excluded regions where the measured \ha/\hb\ ratio is lower than the theoretically expected ratio of 2.86, resulting in a final catalogue size of 15652 nebulae. Our approach to dust extinction here is different from that followed in \citetalias{Congiu2023}, who compared observed fluxes with photoionisation models reddened via a dust screen. We did not use the $E(B-V)$ values from \citetalias{Congiu2023} directly because they were an output of their Bayesian analysis, and we preferred to keep the dataset free from any influence of the models. Using the extinction corrections from \citetalias{Congiu2023}, however, did not lead to significant differences.

In Fig. \ref{fig:data} (lower panel) we show the position of the ionised nebulae selected for analysis in this work in the canonical BPT diagrams, together with the demarcation lines from the literature. It is evident that real \hii\ regions and PNe populate only a fraction of the total space populated by models, while nebulae that have a best-fit classification as SNR extend in a wider section of parameter space than populated by the models.

\section{Methods}
\label{sec:methods}

\subsection{Overview}
\label{sec:overview}

In this section we summarise our machine-learning approach to classification. In this work we compare the performances of a simple NN classifier (trained only on the source domain) and a DANN \citep{Ganin2016}.
A DANN is a `feature-based' domain-adaptation architecture. This class of algorithms derive a set of features which are consistent across both the source and target domains \citep{Ajakan2014}. This domain-invariant representation is then used for classification.

\begin{figure}[ht!]
\centering
\includegraphics[width=0.45\textwidth,trim=0 0 0 0, clip]{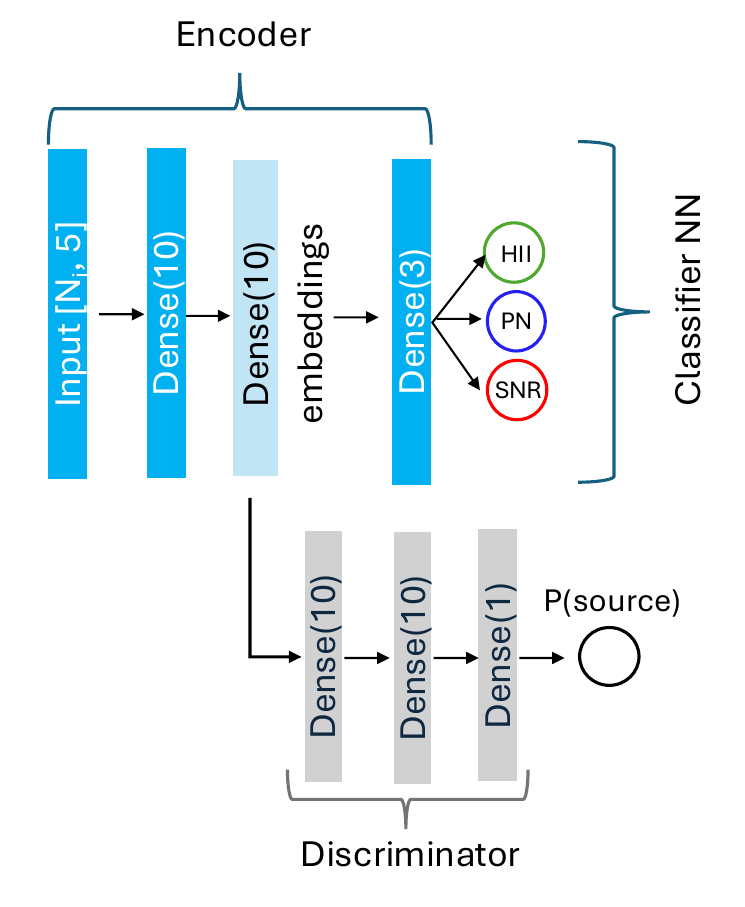}

\caption{Illustration of the network architecture of the simple NN classifier (top row only, blue layers) and the DANN (full figure). Through the cooperation of both the encoder and the discriminator network, the DANN learns domain-invariant features in the embeddings layer (light blue). The DANN is trained adversarially to deceive its discriminator into being unable to distinguish source from target domain instances. The figure also shows the dimensionality of the input (number of instances, $N_i$, by 5) and the number of neurons in each layer (in bracket next to each layer name).}
\label{fig:architecture}
\end{figure}

Specifically, in a DANN both source and target domain instances are used as input to the network. An {\it encoder} network transforms the input data into a latent feature space (embeddings layer in Fig. \ref{fig:architecture}). An additional neural network, called the {\it discriminator}, is passed the embeddings from the encoder. The discriminator's goal is to classify instances as belonging to either the source or the target domain. 

The model is trained adversarially, with the goal of deceiving its discriminator, in a similar fashion to generative adversarial networks (GANs, \citealt{Goodfellow2014}). In particular, for each iteration of the training loop, the DANN is fed both source and target domain instances. The parameters of the encoder network are optimised both for correct classification of source domain instances and for fooling the discriminator. Tricking the discriminator requires the embeddings to minimise any information content that could distinguish the two domains, therefore encoding only features shared by both. This allows the DANN to perform a classification task on domain-invariant features.

The next subsections (Sec. \ref{sec:preprocess}-\ref{sec:DANN}) summarise the algorithmic choices for both networks. Readers less familiar with machine-learning terminology may wish to proceed to the results and astrophysical validation in Sec. \ref{sec:res}.

\subsection{Data preprocessing}
\label{sec:preprocess}

Before feeding the data into our machine-learning pipelines we subdivided the source-domain instances into a training (55\%), validation (20\%), and test (25\%) set. The validation set was used to monitor the model performance during training, perform early stopping, and drive hyper-parameter choices. The test set, on the other hand, was only used to evaluate the model after training. The target domain is unlabelled, so its instances were not subdivided.
All fluxes were normalised to that of \hb\ and we consider the log of the value of the line ratio, so that each instance consists of a vector of five features, namely  log(\oiii$\lambda$5007/\hb), log(\nii$\lambda$6584/\hb), log(\sii$\lambda$6717/\hb), log(\sii$\lambda$6731/\hb), and log(\oi$\lambda$6300/\hb). We did not consider \ha/\hb\ since, after extinction correction, this ratio is fixed to 2.86 in all the observed line fluxes. 
We tested the effect of standardising (subtracting the mean and dividing by the standard deviation) the features before inputting them into the network, but found that adding a batch normalisation layer (see the next section) immediately after the input had comparable performance. We therefore did not standardise the input features. 


\subsection{Architecture of the NN classifier}
\label{sec:ANN}

We designed a simple NN to classify the instances within the source domain. After experimentation, we converged on a simple architecture, comprising an input layer followed by  batch normalisation \citep{Ioffe2015}, two dense hidden layers each with ten neurons, followed each by batch normalisation, and an output dense layer with three neurons, as required for our classification task. We adopt an exponential linear unit (ELU, \citealt{Clevert2015}) activation function for the nodes in the hidden layers, and apply a softmax activation to the output layer to ensure that the network's output represents the probability of categorising each instance into the three target classes.

A sketch of the architecture of the NN is provided in Fig. \ref{fig:architecture}, where the NN corresponds only to the layers in the top part of the diagram. The network was compiled utilising a cross-entropy loss function along with the Adam optimiser \citep{Kingma2014} with a learning rate of $10^{-3}$, and with the initial weights being determined by a He-Normal distribution \citep{He2015}.
The network was built using the flexible \textsc{keras} API with the \textsc{tensorflow} backend.

\subsection{Architecture of the DANN}
\label{sec:DANN}

We constructed a DANN using an encoder network similar to the one in Sec. \ref{sec:ANN} and adding a discriminator and linking it to the last hidden layer of the classifier (the layers whose output constitutes the embeddings). We experimented with various architectures for the discriminator and obtained the best results by mirroring the encoder's architecture. We made one modification to the network with respect to the previous section by adding dropout layers (rate of 0.2) between each two hidden layers. We found that this addition improved performance by reducing overfitting. We also fix the batch size to 128. 
The output layer of the discriminator consists of a single neuron with a sigmoid activation function, as required by the binary classification task of distinguishing between the two domains. We use the implementation of DANN available via the \textsc{adapt} python package.\footnote{\url{https://adapt-python.github.io/adapt/index.html}.}

The total loss function of the DANN is given by 
\begin{equation}
L_{\rm tot} = L_{\rm class} - \lambda~ L_{\rm discr},
\end{equation}
where $L_{\rm class}$ is the classifier cross-entropy loss, $L_{\rm discr}$ is the discriminator binary entropy loss and $\lambda$ is a hyper-parameter indicating the relative weight of the two components. The network was trained with the Adam optimiser and a slower learning rate ($10^{-4}$) than in Sec. \ref{sec:ANN}, as we found in our experiments that a higher learning rate can lead to instabilities. We used a default $\lambda = 1$. We experimented changing $\lambda $ by four orders of magnitude around the default value and found $\lambda = 1$ to lead to the best performance on the target domain.

\subsection{Training the models}

We trained the NN classifier on the labelled source domain, while the DANN was trained on both the labelled source domain and the unlabelled target domain. In particular, the DANN was fed input batches which consisted of half source and half target domain instances.
In all cases, training was halted when the validation loss no longer decreased, employing early stopping and a patience of 20 epochs to prevent overfitting. Each model was trained 20 times with different starting conditions and the results of this ensemble of models were averaged to provide better stability.




\begin{figure}[ht!]
\centering
\includegraphics[width=0.5\textwidth,trim=0 0 30 10, clip]{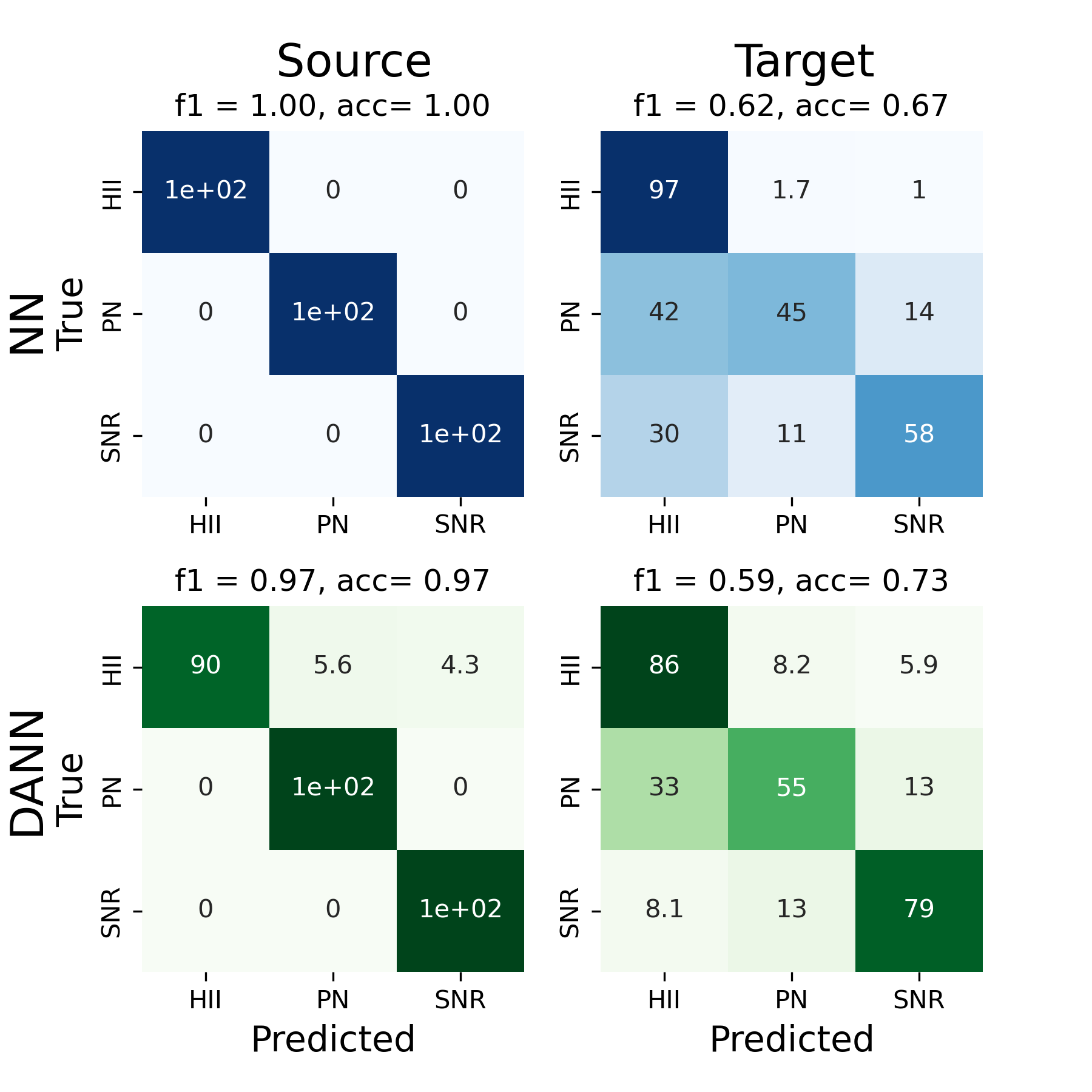}
\caption{
Confusion matrices showing the performance of the NN (top row) and the DANN (bottom row) in classifying \hii\ regions, PNe, and SNRs, evaluated over the test set of the source domain (left column) and the target domain (right column). Values of the f1 score and accuracy (acc) are reported at the top of each panel.
}
\label{fig:conf1}
\end{figure}

\section{Results}
\label{sec:res}

In this section, we present the results of our experiments with domain adaptation and discuss how they compare with those obtained with a simple NN classifier.

\subsection{Benchmarking domain adaptation}
\label{sec:DA}

In Fig. \ref{fig:conf1} we show the confusion matrices obtained when applying the NN and the DANN on the test set of the source domain (left panels) and the target domain (right panels). Each element of the confusion matrix represents the percentage of instances of the input where an instance of a given class is predicted to belong to each of the possible classes. A fully diagonal matrix represents a perfect classifier. In this work, we calculate the confusion matrices considering the average predictions from an ensemble of 20 identical models, trained with different initial conditions.

A simple NN classifier performed extremely well on the test set of the source domain, reaching $\sim100\% $ accuracy. This demonstrates that the three considered classes are sufficiently distinct in the five-dimensional line ratio space considered in this work to always lead to accurate classifications by a simple NN. This result is in general agreement with the analysis of \cite{Rhea2023}, who set up a similar experiment, but considered fewer (three or four) line ratios. \cite{Kang2022} obtained equally good reconstructions using an invertible neural network architecture. 

While we do not have ground truth labels for the target domain (observed data), we nonetheless computed confusion matrices considering the classifications provided by \citetalias{Congiu2023} as ground truth. These classifications are not fully independent of the source domain, since they are based on a Bayesian data-model comparison. They nonetheless correspond to state-of-the-art classifications and we therefore compare against them to benchmark the ability of different models to generalise into the target domain.

The simple NN model did not translate well into the target domain (accuracy 0.67 and f1 score\footnote{The f1 score for a binary classification problem is the harmonic mean of precision and recall. For our multiclass problem we used the average of the f1 scores for each label.} of 0.62). The network particularly struggled to classify SNR and PNe, which were often misclassified as \hii\ regions, while still correctly classifying most true \hii\ regions as such.
The overall disappointing behaviour of the NN classifier on the target domain, despite its excellent performance on the test set of the source domain, clearly demonstrates the existence of a domain shift.

The DANN classifier sacrificed some accuracy in the source domain to improve the performance of the task on the target domain. We trained the DANN 20 times randomly varying the initial weights and show the result of the best model in Fig. \ref{fig:conf1} (bottom panels). This DANN model obtained accuracy of 0.97 in the source domain, and significantly improved in the target domain with respect to the NN classifier, reaching an accuracy score of 0.73. In particular, the DANN improved the behaviour of the classifier for PNe and SNRs in the target domain, which were less commonly confused for \hii\ regions. 


\begin{figure}[t!]
\centering
\includegraphics[width=0.5\textwidth,trim=0 10 0 0, clip]{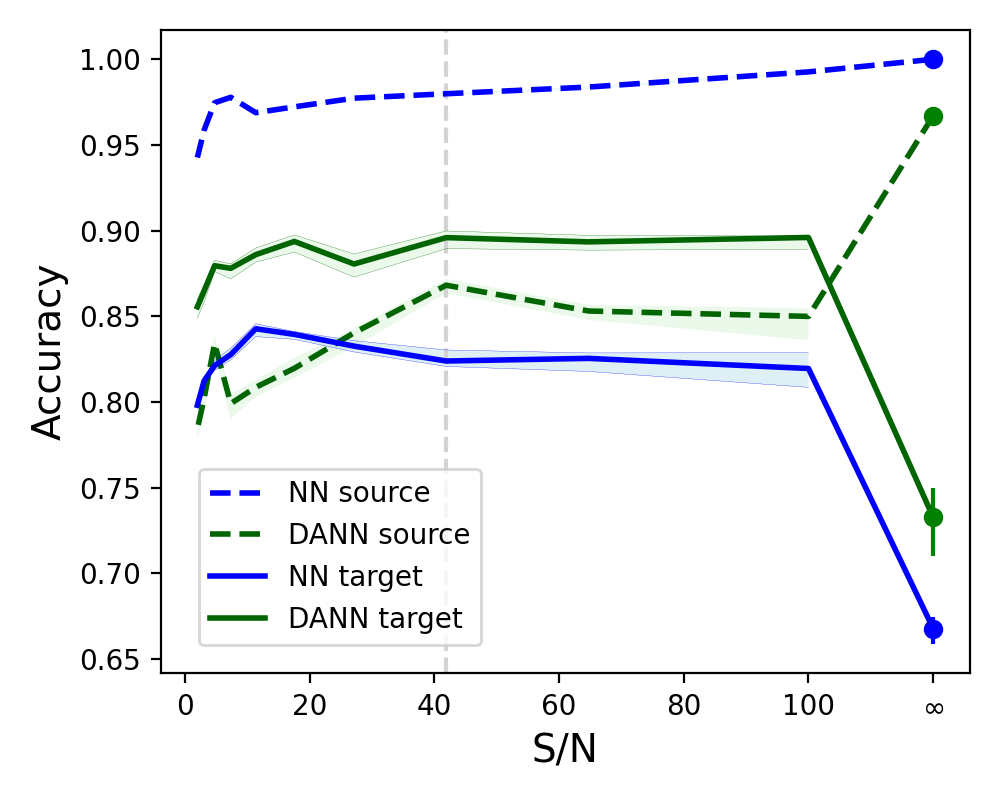}
\caption{
Accuracy on the test set of the source domain and target domain as a function of S/N for the NN (blue) and DANN (green) models. The lines correspond to the average of an ensemble of 20 models and the shaded areas represent the 10$\rm ^{th}$ and 90$\rm ^{th}$ percentiles of the distribution. The accuracy on the target domain for both NN and DANN peaks at  $\mathrm{S/N}\sim40$ (grey dashed line).}
\label{fig:noise_levels}

\centering
\includegraphics[width=0.5\textwidth,trim=0 0 0 0, clip]{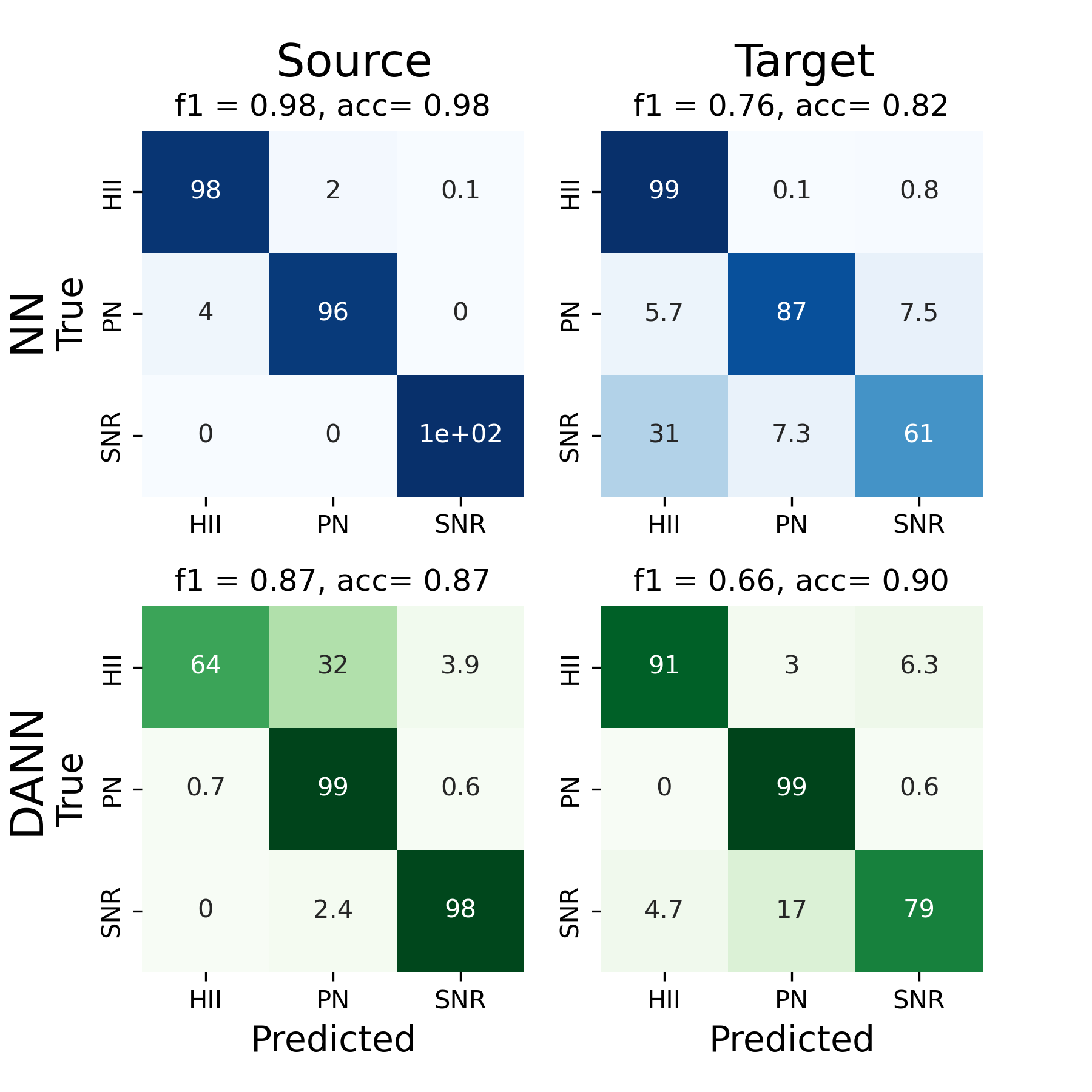}
\caption{Same plots as in Fig. \ref{fig:conf1}, but considering both the NN and the DANN trained over a noise-augmented dataset with S/N$ \sim 40$ on \hb\ in the source domain  (see text in Sec. \ref{sec:noise}). Noise augmentation leads to better performances on the target domain for both the NN and the DANN.}
\label{fig:conf2}
\end{figure}

\begin{figure*}[t]
\centering
\includegraphics[width=1\textwidth,trim=0 0 0 0, clip]{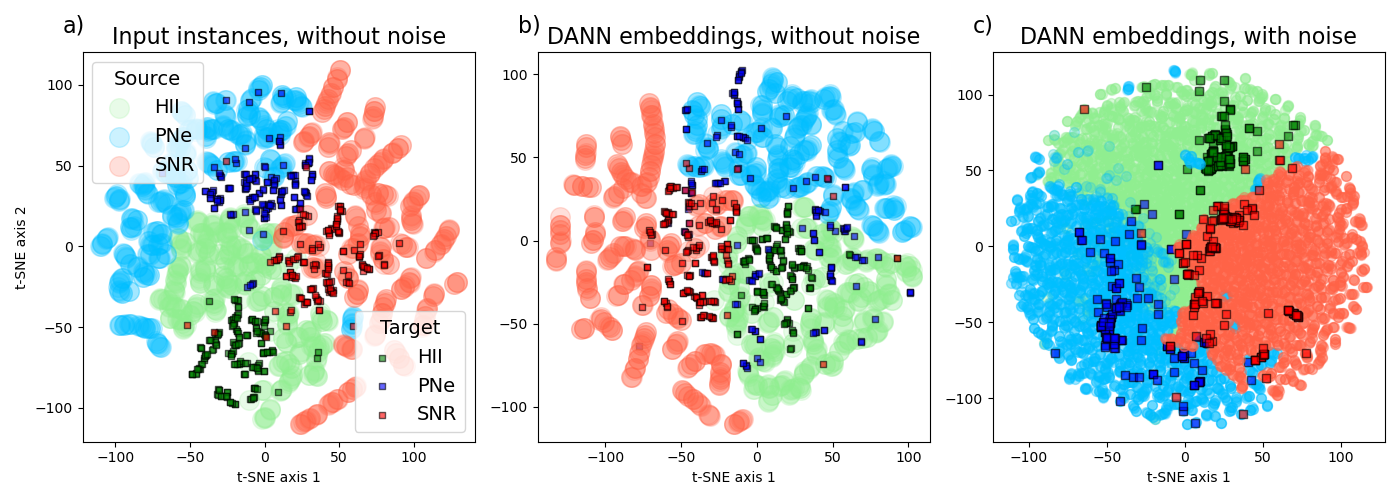}
\caption{
 Two-dimensional representations of the input data and the embeddings from the DANN obtained using t-SNE, illustrating the distribution of instances from both the source and target domains. For ease of visualisation we plot only a random subset of the target domain that homogeneously populates the three classes. The quantities on the axes represent the output of t-SNE and do not have a physical meaning, and they represent a different projection in each panel.
 Light colours (light blue, light red, and light green) represent the three classes in the source domain (\hii\ regions, PNe, and SNRs, respectively), while dark colours (dark blue, dark red, and dark green) represent the same three classes in the target domain. Left: Input vectors show a clear separation between the source and target domains, pointing to the presence of a domain shift. Middle: Embedding vectors from the DANN model show a closer alignment of instances, showcasing the effectiveness of DANN in reducing domain shift. Right: The alignment is further enhanced with noise augmentation in the training set. Dark-coloured outliers in source domain clusters reflect the imperfect accuracy of the DANN model. Their existence is reflected in the non-diagonal confusion matrices discussed in Sec. \ref{sec:DA}.
}
\label{fig:tsne}
\end{figure*}

\subsection{The effect of adding noise}
\label{sec:noise}

We explored the impact of adding noise to the training set of the source domain to determine whether this could enhance the performance of the networks in the target domain. Introducing noise can be considered a simplistic form of domain adaptation, as noise can mimic observational effects, and potentially enforce a form of regularisation, guiding the network to learn more generalisable, domain-invariant features. 

For this test we therefore added Gaussian noise to the training data. In particular, we tested ten noise values logarithmically spaced between 0.01 and 0.5. Since the instances were normalised to an \hb\ flux of one, these noise level correspond to an \hb\ S/N ranging from 0.5 to 100. Other lines would have a different S/N ratio, depending on their brightness with respect to \hb. Such a simplified noise model was therefore assumed to be constant across wavelength.

This approach was not meant to represent a realistic noise distribution, which would be much more complex to simulate, given that \hii\ regions, PNe, and SNRs have different luminosities and S/N ratios in their Balmer line emission. In particular, the S/N ratio on \hb\ for our target sample is $\sim200$ on average, but varies from an average of $300$ for \hii\ regions, to 6 for PNe, and 35 for SNRs using the \citetalias{Congiu2023} labels. A more realistic noise distribution for the different classes in the source domain could be simulated assuming relevant luminosity functions and simulating the expected S/N given the observing parameters of our dataset. This is beyond the scope of our experiment here. However, the noise level does not need to closely match that of our dataset, since domain adaptation should contribute to mitigate the effect of the domain shift even if the noise model is not fully realistic.

We therefore generated a new set of augmented training data consisting of 20160 source instances by adding Gaussian noise, with each instance from the original source domain perturbed ten times. We ensured instances from the same simulation but with different noise realizations were assigned to the same split (train/validation/test) when creating the noise-augmented dataset to avoid potential information leakage. 

We trained the same NN and DANN models using this noise-augmented dataset 20 times for each noise value with different initial random values of the weights. For each ensemble of 20 models we calculated the average prediction probability and used that to compute the accuracy. We obtained errors on this accuracy by bootstrap resampling the distribution 100 times and considering as errors the 10$\rm ^{th}$ and 90$\rm ^{th}$ percentiles of the distribution.
Fig. \ref{fig:noise_levels} shows the resulting average accuracy for the NN and the DANN on the test set of the source domain and on the target domain, with the shaded areas representing the errors. For comparison we also plot the accuracy values for the case with no noise (Sec. \ref{sec:DA}), labelled as S/N$=\infty$. 

The NN was only trained on the source domain and its performance on the test set increases almost monotonically with S/N. The results were virtually identical for all 20 runs. The performance on the NN on the target domain, on the other hand, was much poorer than on the source domain. Moreover the response was not monotonic, peaking at S/N $\sim 10-20$ and decreasing for both lower and higher values of S/N. The worst worst performance of the NN on the target set was obtained with no noise.

For the DANN, the source domain accuracy, as quantified on the test set, also increased with increasing S/N. The target accuracy, however, peaked between S/N$=10-40$, while decreasing for both lower and higher S/N, following a similar trend to the target accuracy of the NN, but shifted to overall higher accuracy values. In the case of no noise, discussed in the previous section, the DANN target performance was lower than its accuracy on the test set, and the results were plagued by instability, as demonstrated by the larger error bar.

These trends match the expectation that adding noise helps the model to generalise, but adding too much noise starts to erase features in the source domain, driving a decrease in accuracy for both source and target domain at $S/N \lesssim 10$. We selected the best model to roughly match the peak of the target accuracy curves, corresponding to S/N $\sim 40$.

Using S/N $\sim 40$, we show the confusion matrices for the performance of the NN and DANN of a representative model (Fig. \ref{fig:conf2}). If we compare directly to the no-noise model, the accuracy of the NN in the target domain increased to 0.82 (compared to 0.67 in the absence of noise). Planetary nebulae are now much better classified, but the model still struggles with SNR. Overall, however, the enhanced performance makes the simple NN classifier even more accurate than the DANN trained in the absence of noise.

The performance of the DANN also increased with the noisy source dataset, reaching an accuracy score in the target domain of 0.90 (compared to 0.73 in the absence of noise), and an f1 score of 0.66. These results show that noise addition can act as a form of regularisation, helping the DANN to learn more robust representations that generalise better across domains.

\begin{figure*}[t]
\centering
\includegraphics[width=1\textwidth,trim=0 0 0 0, clip]{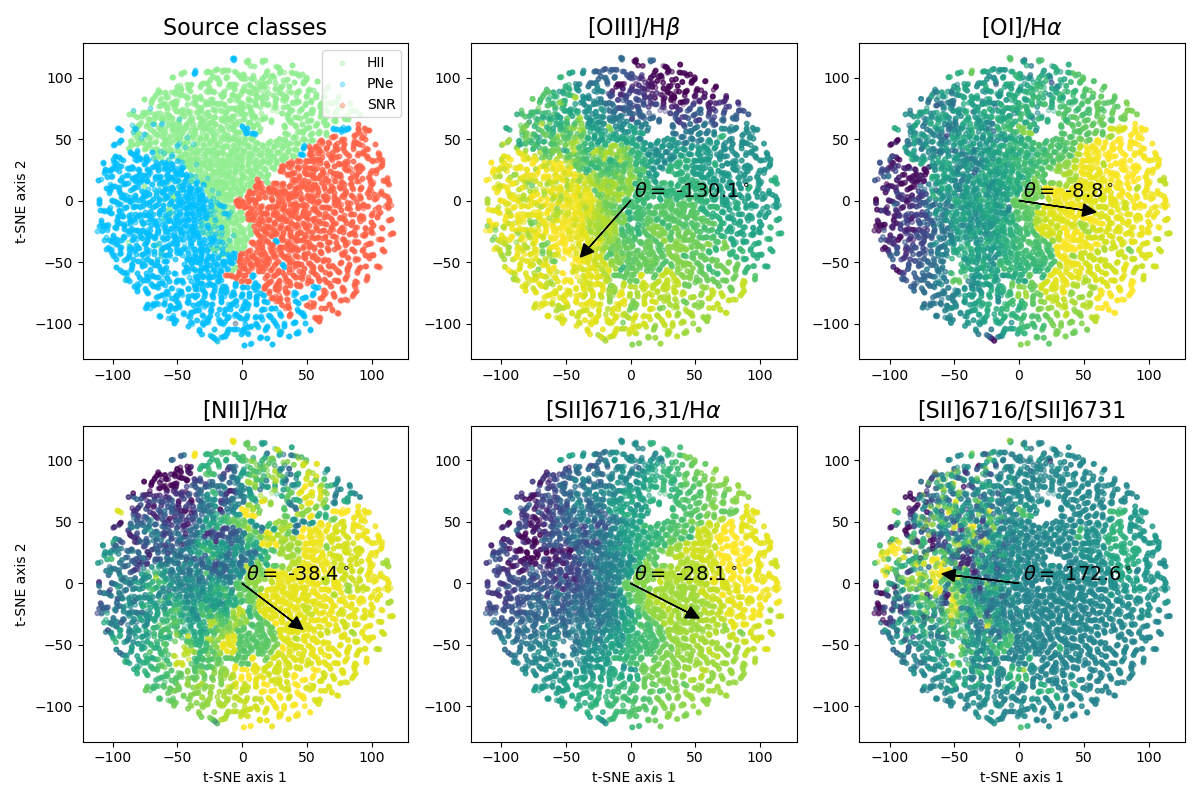}
\caption{
 Two-dimensional representations of the DANN embeddings (a representative model with S/N$\sim$ 40) colour-coded by the source domain classes, and various line ratios (which are either the input features themselves or a combination of input features), shown as the titles of each panel. The colourbar goes from blue (low) to yellow (high) and in each panel is normalised to the 2nd and 98th percentiles of the quantity. The arrow emerging from the origin shows the direction of steepest increase in the colour-coded quantity, measured clockwise from the positive x-axis.
}
\label{fig:interpret_tsne}
\end{figure*}

\subsection{Visualising domain-invariant features}
\label{sec:features}

To visually demonstrate the impact of the domain-adaptation strategy employed, we generated t-SNE (t-Distributed Stochastic Neighbour Embedding; \citealt{VanderMaaten2008}) plots to show the distribution of instances from both the source and target domains. t-SNE is a dimensionality-reduction technique particularly suited for visualising high-dimensional datasets. It works by minimising the divergence between two distributions: a distribution that measures pairwise similarities of the input objects and a distribution that measures pairwise similarities of the corresponding low-dimensional points in the embedding. In our two-dimensional t-SNE, representations points that are similar in high-dimensional space are therefore found nearby in two-dimensional space.

We used t-SNE to visualise the similarities between source and target domain instances, which are five-dimensional vectors of line ratios in Fig. \ref{fig:tsne}a. We use light-colour (light green, light blue, and light red) circles to represent the three classes in the source domain (\hii\ regions, PN, and SNR, respectively), and three dark-colour squares (dark green, dark blue, and dark red) for the same three classes in the target domain. Fig. \ref{fig:tsne}a demonstrates a clear separation between the source and target domains. In particular the \hii\ regions and PNe target domain instances fall within gaps of the parameter space not covered by source domain instances, indicating a domain shift, and the overlap is only partial for SNRs.

We repeated this analysis using the embedding vectors from the DANN model, which are ten-dimensional, in Fig. \ref{fig:tsne}b. As discussed in the previous sections, the DANN model aligns the distributions of the source and target domain features in the embeddings space. As shown in Fig \ref{fig:tsne}b, a closer alignment between the source and target domain instances is observed, demonstrated by the overlap of light and dark points within the same clusters, indicating the effectiveness of DANN in reducing the domain shift.

Introducing noise in the training set enhanced the domain-adaptation capability of the DANN, inducing a better alignment of the embeddings from both domains, as shown in Fig. \ref{fig:tsne}c. For this example we show a representative S/N $\sim$ 40 DANN model and demonstrate that light (source) and dark (target domain) data points now largely occupy the same regions of the two-dimensional space. Nonetheless, a slight `contamination' of classes is observed, evidenced by dark-coloured outliers situated within the source domain clusters. In fact, we find that these instances correspond to cases where the DANN does not predict the correct class. 

This visual analysis provides a tangible demonstration of how domain adaptation, as implemented by a DANN and further enhanced by noise augmentation, aids in aligning the distributions of the source and target domains, thus addressing the domain-shift challenge.

\subsection{Interpreting the DANN embeddings}
\label{sec:interpret}

We used the dimensionality-reduced output of the DANN embedding layer discussed in the previous section to gain qualitative insights into what information the model has encoded. Visualising the low-dimensional output of key layers in a neural network is a well-established interpretability technique, which has also been used in an unsupervised fashion to learn about the underlying structure in the data \citep{Portillo2020, Sarmiento2021, Baron2024, Ginolfi2024}.

In Fig. \ref{fig:interpret_tsne} we present the dimensionality-reduced DANN embeddings of the source domain colour-coded by the line ratios which are used as input to our model or combinations thereof. To ease interpretability, in the case of the sulphur line doublet we consider their sum  as (\sii $\lambda$6717+ \sii$\lambda$6731)/\ha\ and their density-sensitive ratio (\sii$\lambda$6717/\sii$\lambda$6731) instead. For reference we also plot in the top-left panel the source domain classes (same as Fig. \ref{fig:tsne}c). 

We observe coherent gradients in key line ratios across the two-dimensional space considered. We quantified the direction of these gradients by  calculating the partial correlation coefficients between the quantities on the x and y-axis ($X$ and $Y$) and the quantity we use for colour-coding ($Z$). In particular we followed \cite{Baker2023} and determined the angle of maximum gradient ($\theta$) by calculating
\begin{equation}
\tan(\theta)  = \frac{\rho_{YZ|X}}{\rho_{XZ|Y}},
\end{equation}
where $\rho_{YZ|X}$ is the partial correlation coefficient between $X$ and $Y$ controlling for $X$. In each panel of Fig. \ref{fig:interpret_tsne} the arrows emerging from the origin show these angles, which are measured clockwise from the positive horizontal direction and evidently point towards the increasing gradient of the colour-coded quantity.

We found that the \oiii/\hb\ ratio increases towards the bottom-right ($\theta = -130.1 ^\circ$) and high values are therefore mostly associated with PNe, as expected. The low-ionisation line ratios (\oi/\ha, \nii/\ha\ and \sii/\ha) increase fastest towards the bottom-right, with \nii/\ha\ ($\theta = -38.4^\circ$) being more similar to \sii/\ha\ ($\theta = -28.1^\circ$) than \oi/\ha\ ($\theta = -8.8^\circ$). These differences can mostly be attributed to the fact that the lowest line ratios in both \nii/\ha\ and \sii/\ha\ are found at the boundary between the \hii\ and PNe classes, while the lowest \oi/\ha\ occur in PNe only. Finally the density-sensitive \sii$\lambda$6717/\sii$\lambda$6731 ratio only shows some enhancement for the high-density \hii\ region and PNe models but no overall gradient.
Interestingly, the direction of maximum change for \oiii/\hb\ and \nii/\ha\ are nearly perpendicular (difference of 92$^\circ$), highlighting the ability of these two line ratios to efficiently span the two-dimensional space considered and therefore confirming the usefulness of the \nii-BPT in performing classifications of ionised nebulae.

\subsection{Astrophysical validation with specialised catalogues}
\label{sec:catalog_compare}

Because of the substantial difficulties in addressing the general problem of detection and classification of nebulae, some authors have produced catalogues aimed at a specific class of objects, using selection criteria fine-turned for that particular class. \cite{Scheuermann2022}, for example, produced a catalogue of PNe from the PHANGS-MUSE dataset specifically selecting unresolved \oiii\ sources and applying additional cuts in line-ratio diagrams. \cite{Li2024} generated a catalogue of SNRs from the same dataset. They selected candidate regions based on a source-finding algorithm using the line ratios \sii/\ha\ and \oi/\ha\ after correcting for diffuse ionised gas emission \citep{Belfiore2022}, and considered additional kinematic (velocity dispersion), and line-ratio information. 

\begin{figure}[ht!]
\centering
\includegraphics[width=0.5\textwidth,trim=0 0 0 0, clip]{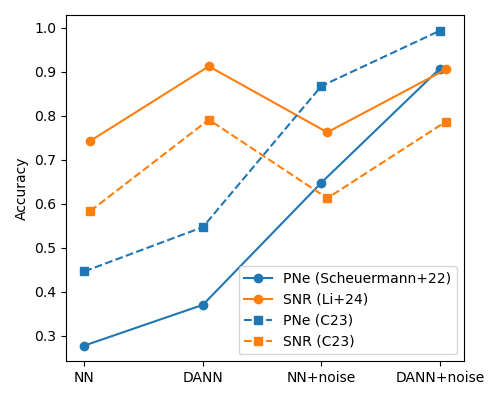}
\caption{
Accuracy of the NN and DANN models (trained both on the noise-free and the noisy source domain) on the target domain, taking as ground truth the classifications of \citetalias{Congiu2023} (the default in this work), the PNe catalogue of \cite{Scheuermann2022}, and the SNR catalogue of \cite{Li2024}, respectively. The DANN+noise model reaches the highest levels of accuracy for both PNe and SNRs across all catalogues.
}
\label{fig:compare_catalogs}
\end{figure}

We tested our models on the PNe catalogue of \cite{Scheuermann2022} and the SNRs catalogue of \cite{Li2024}. To do so, we cross-matched these catalogues to the one of \citetalias{Congiu2023}, matching objects with a distance of $< 0.4 ''$. We then pre-processed these catalogues in the same way as the \citetalias{Congiu2023} data. We show the resulting (binary) accuracy results in Fig. \ref{fig:compare_catalogs}. We did not re-train the models but simply consider the accuracy of the predictions between the class of interest and all other classes.

Fig. \ref{fig:compare_catalogs} shows that both domain adaptation and noise addition lead to improvements in classification accuracy, although in different ways for different classes. For PNe, for example, both the addition of noise and the DANN contribute substantially to the increase in accuracy. The performance on the \cite{Scheuermann2022} catalogue remains slightly lower than on the \citetalias{Congiu2023} catalogue, however. For SNRs, on the other hand, all models perform better on the  \cite{Li2024} catalogue than on \citetalias{Congiu2023}, and the addition of noise does not improve performance substantially. For the DANN+noise case, the model reached an accuracy $>$ 79\% for all the catalogues of both PNe and SNRs.

\section{Summary and conclusions}
\label{sec:conclusions}

Several astrophysical science cases require the classification of objects based on a comparison between imperfect models and observational data. This class of problems can be tackled by machine-learning methods using domain adaptation, where a model is trained on labelled source-domain data (the theory) with the goal of optimising its performance on the unlabelled target domain (the observational data). In this work we presented a proof-of-concept application of domain-adaptation techniques to the challenge of classifying ionised nebulae into \hii\ regions, planetary nebulae and supernova remnants based on their line ratios. Our source domain is represented by a set of libraries of photoionisation and shock simulations from the literature \citep{Perez-Montero2014, Delgado-Inglada2014, Allen2008}. We chose as target domain the line ratios measured in the sample of nearby PHANGS--MUSE galaxies by \citetalias{Congiu2023}. 

We showed that a primary obstacle to an accurate classification is the domain shift that occurs when applying existing photoionisation simulations to real-world astronomical data. To address this, we employed a domain-adversarial neural
network (DANN) framework and compared its performance to that of a vanilla feed-forward network classifier (NN). Our main conclusions are described below.


\begin{itemize}
    \item The NN classifier showed excellent performance on the source domain (100\% accuracy on the test set), but considering the classifications carried out in \citetalias{Congiu2023} as the ground truth, its performance on the target domain was severely affected by domain shift (67\% accuracy and f1 score of 0.62).
    
     \item The DANN demonstrated superior performance over a classical NN classifier on the target domain (reaching 73\% accuracy), indicating the effectiveness of domain-adaptation techniques in countering the domain-shift challenge. This comes at the cost of lower accuracy on the source domain.

    \item Adding noise to the input models acts as a form of regularisation, enhancing the performance of both the NN classifier and the DANN on the target domain. In particular, the best performance on the target domain was obtained with S/N $=10-40$, while very high noise leads to decreased classification accuracy, presumably as information is erased from the weaker features in the source domain.

    \item By combining the use of a DANN and artificial noise injection in the models we achieved a notable improvement in classification accuracy, with our DANN best model reaching a classification accuracy of 90\% (f1 score of 0.66) on the target domain. 

    \item We demonstrated that the DANN leads to closer alignment of the source and target domain instances by projecting their embeddings onto a two-dimensional parameter space via t-SNE. We also showed using this low-dimensional representation that the information learnt by the DANN can be interpreted in terms of input line ratios.

    \item We applied our best DANN+noise model to specialised PNe \citep{Scheuermann2022} and SNRs \citep{Li2024} catalogues and found that both noise addition and domain adaptation lead to substantial improvement in the accuracy of classifying nebulae of these two classes also for these specialised catalogues.
\end{itemize}

Even for the relatively simple problem we have studied in this work, the combination of domain adaptation and noise injection in the models resulted in an increase in accuracy in the target domain of 23\% over the naive application of a NN on the noise-free models. While the combination of domain adaptation and noise injection led to the best solution, domain adaptation on the noise-free models already represented a substantial improvement over the use of the NN on noise-free data. This highlights the potential of domain-adaptation techniques to a wide variety of astrophysical problems, where no simple recipe may exist to mimic observational realism. As astrophysics proceeds into the era of big data and large survey projects, domain adaptation is likely to become a fundamental component of ML-powered data analysis pipelines.

\begin{acknowledgements}

We thank the referee for the thoughtful comments which improved the quality of the paper.

F.B thanks Ricarda Sontag for stimulating discussions that have improved the quality of this work.

This work has been carried out as part of the PHANGS
collaboration. Based on observations from the PHANGS-MUSE programme, collected at the European Southern Observatory under ESO programmes 094.C-0623 (PI: Kreckel), 095.C-0473,  098.C-0484 (PI: Blanc), 1100.B-0651 (PHANGS-MUSE; PI: Schinnerer), as well as 094.B-0321 (MAGNUM; PI: Marconi), 099.B-0242, 0100.B-0116, 098.B-0551 (MAD; PI: Carollo) and 097.B-0640 (TIMER; PI: Gadotti).

FB acknowledges support from the INAF Fundamental Astrophysics programme 2022. G.A.B. acknowledge support from the ANID Basal project FB210003.

RSK acknowledges financial support from the European Research Council via the ERC Synergy Grant ``ECOGAL'' (project ID 855130),  from the German Excellence Strategy via the Heidelberg Cluster of Excellence (EXC 2181 - 390900948) ``STRUCTURES'', and from the German Ministry for Economic Affairs and Climate Action in project ``MAINN'' (funding ID 50OO2206). 

RSK is grateful for computing resources provided by the Ministry of Science, Research and the Arts (MWK) of the State of Baden-W\"{u}rttemberg through bwHPC and the German Science Foundation (DFG) through grants INST 35/1134-1 FUGG and 35/1597-1 FUGG, and also for data storage at SDShd funded through grants INST 35/1314-1 FUGG and INST 35/1503-1 FUGG.

RSK also thanks the Harvard-Smithsonian Center for Astrophysics and the Radcliffe Institute for Advanced Studies for their hospitality during his sabbatical, and the 2024/25 Class of Radcliffe Fellows for highly interesting and stimulating discussions.
MB gratefully acknowledges support from the ANID BASAL project FB210003 and from the FONDECYT regular grant 1211000.

This work was supported by the French government through the France 2030 investment plan managed by the National Research Agency (ANR), as part of the Initiative of Excellence of Université Côte d’Azur under reference number ANR-15-IDEX-01.

\end{acknowledgements}


%
\bibliographystyle{aa} 
\bibliography{library9} 

\begin{thebibliography}{65}
\expandafter\ifx\csname natexlab\endcsname\relax\def\natexlab#1{#1}\fi

\bibitem[{Ajakan {et~al.}(2014)Ajakan, Germain, Larochelle, Laviolette, \&
  Marchand}]{Ajakan2014}
Ajakan, H., Germain, P., Larochelle, H., Laviolette, F., \& Marchand, M. 2014,
  arXiv, arXiv:1412.4446

\bibitem[{Allen {et~al.}(2008)Allen, Groves, Dopita, Sutherland, \&
  Kewley}]{Allen2008}
Allen, M., Groves, B., Dopita, M., Sutherland, R., \& Kewley, L. 2008, ApJS,
  178, 20

\bibitem[{Amayo {et~al.}(2021)Amayo, Delgado-Inglada, \& Stasinska}]{Amayo2021}
Amayo, A., Delgado-Inglada, G., \& Stasinska, G. 2021, MNRAS, 505, 2361

\bibitem[{Baker {et~al.}(2023)Baker, Maiolino, Belfiore, Curti, Bluck, Lin,
  Ellison, Thorp, Pan, Osservatorio, Fermi, \& Florence}]{Baker2023}
Baker, W.~M., Maiolino, R., Belfiore, F., {et~al.} 2023, MNRAS, 519, 1149

\bibitem[{Baldwin {et~al.}(1981)Baldwin, Phillips, \& Terlevich}]{Baldwin1981}
Baldwin, J.~A., Phillips, M.~M., \& Terlevich, R. 1981, PASP, 93, 5

\bibitem[{Barnes {et~al.}(2020)Barnes, Longmore, Dale, Krumholz, Kruijssen, \&
  Bigiel}]{Barnes2020}
Barnes, A.~T., Longmore, S.~N., Dale, J.~E., {et~al.} 2020, MNRAS, 498, 4906

\bibitem[{{Baron} {et~al.}(2024){Baron}, {Sandstrom}, {Rosolowsky}, {Egorov},
  {Klessen}, {Leroy}, {Boquien}, {Schinnerer}, {Belfiore}, {Groves},
  {Chastenet}, {Dale}, {Blanc}, {M{\'e}ndez-Delgado}, {Koch}, {Grasha},
  {Chevance}, {Thilker}, {Colombo}, {Williams}, {Pathak}, {Sutter}, {Brown},
  {Wu}, {Peek}, {Emsellem}, {Larson}, \& {Neumann}}]{Baron2024}
{Baron}, D., {Sandstrom}, K.~M., {Rosolowsky}, E., {et~al.} 2024, ApJ, 968, 24

\bibitem[{Belfiore {et~al.}(2016)Belfiore, Maiolino, \&
  Bothwell}]{Belfiore2016}
Belfiore, F., Maiolino, R., \& Bothwell, M. 2016, MNRAS, 455, 1218

\bibitem[{Belfiore {et~al.}(2022)Belfiore, Santoro, Groves, Schinnerer,
  Kreckel, Glover, Klessen, Emsellem, Blanc, Congiu, Barnes, Boquien, Chevance,
  Dale, {Diederik Kruijssen}, Leroy, Pan, Pessa, Schruba, \&
  Williams}]{Belfiore2022}
Belfiore, F., Santoro, F., Groves, B., {et~al.} 2022, A{\&}A, 659, A26

\bibitem[{Binette {et~al.}(2012)Binette, Matadamas, H{\"{a}}gele, Nicholls, \&
  Magris}]{Binette2012}
Binette, L., Matadamas, R., H{\"{a}}gele, G.~F., Nicholls, D.~C., \& Magris,
  G.~C. 2012, A{\&}A, 547, A29

\bibitem[{{Bisbas} {et~al.}(2009){Bisbas}, {W{\"u}nsch}, {Whitworth}, \&
  {Hubber}}]{Bisbas2009}
{Bisbas}, T.~G., {W{\"u}nsch}, R., {Whitworth}, A.~P., \& {Hubber}, D.~A. 2009,
  A\&A, 497, 649

\bibitem[{Bottrell {et~al.}(2019)Bottrell, Hani, Teimoorinia, Ellison, Moreno,
  Torrey, Hayward, Thorp, Simard, \& Hernquist}]{Bottrell2019}
Bottrell, C., Hani, M.~H., Teimoorinia, H., {et~al.} 2019, MNRAS, 490, 5390

\bibitem[{Byler {et~al.}(2019)Byler, Dalcanton, Conroy, Johnson, Choi, Dotter,
  \& Rosenfield}]{Byler2019}
Byler, N., Dalcanton, J.~J., Conroy, C., {et~al.} 2019, AJ, 158, 2

\bibitem[{Chevance {et~al.}(2023)Chevance, Krumholz, McLeod, Ostriker,
  Rosolowsky, \& Sternberg}]{Chevance2023}
Chevance, M., Krumholz, M., McLeod, A., {et~al.} 2023, in Astronomical Society
  of the Pacific Conference Series, Vol. 534, Protostars Planets VII, ed.
  S.~Inutsuka, Y.~Aikawa, T.~Muto, K.~Tomida, \& M.~Tamura, 1

\bibitem[{{Cid Fernandes} {et~al.}(2011){Cid Fernandes}, Stasi{\'{n}}ska,
  Mateus, \& {Vale Asari}}]{CidFernandes2011}
{Cid Fernandes}, R., Stasi{\'{n}}ska, G., Mateus, A., \& {Vale Asari}, N. 2011,
  MNRAS, 413, 1687

\bibitem[{Ciprijanovic {et~al.}(2021)Ciprijanovic, Kafkes, Downey, Jenkins,
  Perdue, Madireddy, Johnston, Snyder, \& Nord}]{Ciprijanovic2021}
Ciprijanovic, A., Kafkes, D., Downey, K., {et~al.} 2021, MNRAS, 506, 677

\bibitem[{{Clevert} {et~al.}(2015){Clevert}, {Unterthiner}, \&
  {Hochreiter}}]{Clevert2015}
{Clevert}, D.-A., {Unterthiner}, T., \& {Hochreiter}, S. 2015, arXiv,
  arXiv:1511.07289

\bibitem[{Congiu {et~al.}(2023)Congiu, Blanc, Belfiore, Santoro, Scheuermann,
  Kreckel, Emsellem, Groves, Pan, Bigiel, Dale, Glover, Grasha, Egorov, Leroy,
  Schinnerer, Watkins, \& Williams}]{Congiu2023}
Congiu, E., Blanc, G.~A., Belfiore, F., {et~al.} 2023, A{\&}A, 672, A148

\bibitem[{Delgado-Inglada {et~al.}(2014)Delgado-Inglada, Morisset, \&
  Stasi{\'{n}}ska}]{Delgado-Inglada2014}
Delgado-Inglada, G., Morisset, C., \& Stasi{\'{n}}ska, G. 2014, MNRAS, 440, 536

\bibitem[{Emsellem {et~al.}(2022)Emsellem, Schinnerer, Santoro, Belfiore,
  Pessa, McElroy, Blanc, Congiu, Groves, Ho, Kreckel, Razza, Sanchez-Blazquez,
  Egorov, Faesi, Klessen, Leroy, Meidt, Querejeta, Rosolowsky, Scheuermann,
  Anand, Barnes, Bes, Bigiel, Boquien, Cao, Chevance, Dale, Eibensteiner,
  Glover, Grasha, Henshaw, Hughes, Koch, Kruijssen, Lee, Liu, Pan, Pety, Saito,
  Sandstrom, Schruba, Sun, Thilker, Usero, Watkins, \& Williams}]{Emsellem2022}
Emsellem, E., Schinnerer, E., Santoro, F., {et~al.} 2022, A{\&}A, 659, A191

\bibitem[{Ferland {et~al.}(1994)Ferland, Binette, Contini, Harrington, Kallman,
  Netzer, P{\'{e}}quignot, Raymond, Rubin, Shields, Sutherland, \&
  Viegas}]{Ferland1994}
Ferland, G., Binette, L., Contini, M., {et~al.} 1994, Anal. Emiss. Lines, 83

\bibitem[{Ferland {et~al.}(2017)Ferland, Chatzikos, Guzm{\'{a}}n, Lykins, {Van
  Hoof}, Williams, Abel, Badnell, Keenan, Porter, \& Stancil}]{Ferland2017}
Ferland, G.~J., Chatzikos, M., Guzm{\'{a}}n, F., {et~al.} 2017, RMxAA, 53, 385

\bibitem[{Ganin {et~al.}(2016)Ganin, Ustinova, Ajakan, Germain, Larochlle,
  Laviolette, Marchand, \& Lempitsky}]{Ganin2016}
Ganin, Y., Ustinova, E., Ajakan, H., {et~al.} 2016, J. Mach. Learn. Res., 17, 1

\bibitem[{{Ginolfi} {et~al.}(2024){Ginolfi}, {Mannucci}, {Belfiore}, {Marconi},
  {Boardman}, {Pozzetti}, {Bolzonella}, {Di Teodoro}, {Cresci}, {Wild},
  {Rodrigues}, {Maiolino}, {Cirasuolo}, \& {Oliva}}]{Ginolfi2024}
{Ginolfi}, M., {Mannucci}, F., {Belfiore}, F., {et~al.} 2024, arXiv,
  arXiv:2410.22420

\bibitem[{{Goodfellow} {et~al.}(2014){Goodfellow}, {Pouget-Abadie}, {Mirza},
  {Xu}, {Warde-Farley}, {Ozair}, {Courville}, \& {Bengio}}]{Goodfellow2014}
{Goodfellow}, I.~J., {Pouget-Abadie}, J., {Mirza}, M., {et~al.} 2014, arXiv,
  arXiv:1406.2661

\bibitem[{He {et~al.}(2015)He, Zhang, Ren, \& Sun}]{He2015}
He, K., Zhang, X., Ren, S., \& Sun, J. 2015, arXiv, arXiv:1502.01852

\bibitem[{Ioffe \& Szegedy(2015)}]{Ioffe2015}
Ioffe, S. \& Szegedy, C. 2015, arXiv, arXiv:1502.03167

\bibitem[{Jin {et~al.}(2022)Jin, Kewley, \& Sutherland}]{Jin2022a}
Jin, Y., Kewley, L.~J., \& Sutherland, R.~S. 2022, ApJ, 934, L8

\bibitem[{{Kang} {et~al.}(2023){Kang}, {Klessen}, {Ksoll}, {Ardizzone},
  {Koethe}, \& {Glover}}]{Kang2023}
{Kang}, D.~E., {Klessen}, R.~S., {Ksoll}, V.~F., {et~al.} 2023, MNRAS, 520,
  4981

\bibitem[{Kang {et~al.}(2022)Kang, Pellegrini, Ardizzone, Klessen, Koethe,
  Glover, \& Ksoll}]{Kang2022}
Kang, D.~E., Pellegrini, E.~W., Ardizzone, L., {et~al.} 2022, MNRAS, 512, 617

\bibitem[{Kauffmann {et~al.}(2003)Kauffmann, Heckman, Tremonti, Brinchmann,
  Charlot, White, Ridgway, Brinkmann, Fukugita, Hall, Ivezi{\'{c}}, Richards,
  \& Schneider}]{Kauffmann2003a}
Kauffmann, G., Heckman, T.~M., Tremonti, C., {et~al.} 2003, MNRAS, 346, 1055

\bibitem[{Kewley {et~al.}(2001)Kewley, Dopita, Sutherland, Heisler, \&
  Trevena}]{Kewley2001}
Kewley, L.~J., Dopita, M.~A., Sutherland, R.~S., Heisler, C.~A., \& Trevena, J.
  2001, ApJ, 556, 121

\bibitem[{Kewley {et~al.}(2006)Kewley, Groves, Kauffmann, \&
  Heckman}]{Kewley2006}
Kewley, L.~J., Groves, B., Kauffmann, G., \& Heckman, T. 2006, MNRAS, 372, 961

\bibitem[{Kewley {et~al.}(2019)Kewley, Nicholls, Sutherland, Rigby, Acharya,
  Dopita, \& Bayliss}]{Kewley2019}
Kewley, L.~J., Nicholls, D.~C., Sutherland, R., {et~al.} 2019, ApJ, 880, 16

\bibitem[{Kingma \& Ba(2014)}]{Kingma2014}
Kingma, D.~P. \& Ba, J. 2014, arXiv, arXiv:1412.6980

\bibitem[{{Kopsacheili} {et~al.}(2020){Kopsacheili}, {Zezas}, \&
  {Leonidaki}}]{Kopsacheili2020}
{Kopsacheili}, M., {Zezas}, A., \& {Leonidaki}, I. 2020, MNRAS, 491, 889

\bibitem[{Law {et~al.}(2021)Law, Gordon, \& Misselt}]{Law2021}
Law, K.-h., Gordon, K.~D., \& Misselt, K.~A. 2021, ApJ, 920, 96

\bibitem[{{Li} {et~al.}(2024){Li}, {Kreckel}, {Sarbadhicary}, {Egorov},
  {Groves}, {Long}, {Congiu}, {Belfiore}, {Glover}, {Barnes}, {Bigiel},
  {Blanc}, {Grasha}, {Klessen}, {Leroy}, {Lopez}, {M{\'e}ndez-Delgado},
  {Neumann}, {Schinnerer}, {Williams}, \& {collaborators}}]{Li2024}
{Li}, J., {Kreckel}, K., {Sarbadhicary}, S., {et~al.} 2024, arXiv,
  arXiv:2405.08974

\bibitem[{Lopez {et~al.}(2014)Lopez, Krumholz, Bolatto, Prochaska,
  Ramirez-Ruiz, \& Castro}]{Lopez2014}
Lopez, L.~A., Krumholz, M.~R., Bolatto, A.~D., {et~al.} 2014, ApJ, 795, 121

\bibitem[{Maiolino \& Mannucci(2019)}]{Maiolino2019}
Maiolino, R. \& Mannucci, F. 2019, A{\&}AR, 27, 3

\bibitem[{McLeod {et~al.}(2021)McLeod, Ali, Chevance, {Della Bruna}, Schruba,
  Stevance, Adamo, Kruijssen, Longmore, Weisz, \& Zeidler}]{McLeod2021}
McLeod, A.~F., Ali, A.~A., Chevance, M., {et~al.} 2021, MNRAS, 508, 5425

\bibitem[{M{\'{e}}ndez-Delgado {et~al.}(2023)M{\'{e}}ndez-Delgado, Esteban,
  Garc$\backslash$'$\backslash$ia-Rojas, Kreckel, \&
  Peimbert}]{Mendez-Delgado2023}
M{\'{e}}ndez-Delgado, J.~E., Esteban, C.,
  Garc$\backslash$'$\backslash$ia-Rojas, J., Kreckel, K., \& Peimbert, M. 2023,
  Natur, 618, 249

\bibitem[{{M{\'e}ndez-Delgado} {et~al.}(2023){M{\'e}ndez-Delgado}, {Esteban},
  {Garc{\'\i}a-Rojas}, {Arellano-C{\'o}rdova}, {Kreckel}, {G{\'o}mez-Llanos},
  {Egorov}, {Peimbert}, \& {Orte-Garc{\'\i}a}}]{Mendez-Delgado2023a}
{M{\'e}ndez-Delgado}, J.~E., {Esteban}, C., {Garc{\'\i}a-Rojas}, J., {et~al.}
  2023, MNRAS, 523, 2952

\bibitem[{Micheva {et~al.}(2022)Micheva, Roth, Weilbacher, Morisset, Castro,
  {Monreal Ibero}, Soemitro, Maseda, Steinmetz, \& Brinchmann}]{Micheva2022}
Micheva, G., Roth, M.~M., Weilbacher, P.~M., {et~al.} 2022, A{\&}A, 668, 1

\bibitem[{Mingozzi {et~al.}(2019)Mingozzi, Cresci, Venturi, Marconi, Mannucci,
  Perna, Belfiore, Carniani, Balmaverde, Brusa, Cicone, Feruglio, Gallazzi,
  Mainieri, Maiolino, Nagao, Nardini, Sani, Tozzi, \& Zibetti}]{Mingozzi2019}
Mingozzi, M., Cresci, G., Venturi, G., {et~al.} 2019, A{\&}A, 622, A146

\bibitem[{Morisset {et~al.}(2015)Morisset, Delgado-Inglada, \&
  Flores-Fajardo}]{Morisset2015}
Morisset, C., Delgado-Inglada, G., \& Flores-Fajardo, N. 2015, RMxAA, 51, 101

\bibitem[{O'Donnell(1994)}]{O'Donnell1994}
O'Donnell, J.~E. 1994, ApJ, 422, 158

\bibitem[{Pearson {et~al.}(2019)Pearson, Wang, Trayford, Petrillo, \& {Van Der
  Tak}}]{Pearson2019}
Pearson, W.~J., Wang, L., Trayford, J.~W., Petrillo, C.~E., \& {Van Der Tak},
  F.~F. 2019, A{\&}A, 626, 1

\bibitem[{Peimbert {et~al.}(2017)Peimbert, Peimbert, \&
  Delgado-Inglada}]{Peimbert2017}
Peimbert, M., Peimbert, A., \& Delgado-Inglada, G. 2017, PASP, 129, 82001

\bibitem[{Pellegrini {et~al.}(2020{\natexlab{a}})Pellegrini, Rahner, Reissl,
  Glover, Klessen, Rousseau-Nepton, \& Herrera-Camus}]{Pellegrini2020a}
Pellegrini, E.~W., Rahner, D., Reissl, S., {et~al.} 2020{\natexlab{a}}, MNRAS,
  496, 339

\bibitem[{Pellegrini {et~al.}(2020{\natexlab{b}})Pellegrini, Reissl, Rahner,
  Klessen, Glover, Pakmor, Herrera-Camus, \& Grand}]{Pellegrini2020}
Pellegrini, E.~W., Reissl, S., Rahner, D., {et~al.} 2020{\natexlab{b}}, MNRAS,
  498, 3193

\bibitem[{Perez-Montero(2014)}]{Perez-Montero2014}
Perez-Montero, E. 2014, MNRAS, 441, 2663

\bibitem[{{Portillo} {et~al.}(2020){Portillo}, {Parejko}, {Vergara}, \&
  {Connolly}}]{Portillo2020}
{Portillo}, S. K.~N., {Parejko}, J.~K., {Vergara}, J.~R., \& {Connolly}, A.~J.
  2020, AJ, 160, 45

\bibitem[{{Rhea} {et~al.}(2023){Rhea}, {Rousseau-Nepton}, {Moumen}, {Prunet},
  {Hlavacek-Larrondo}, {Grasha}, {Robert}, {Morisset}, {Stasinska},
  {Vale-Asari}, {Giroux}, {McLeod}, {Gendron-Marsolais}, {Wang}, {Lyman}, \&
  {Chemin}}]{Rhea2023}
{Rhea}, C.~L., {Rousseau-Nepton}, L., {Moumen}, I., {et~al.} 2023, RASTI, 2,
  345

\bibitem[{Roth {et~al.}(2018)Roth, Sandin, Kamann, Husser, Weilbacher,
  Monreal-Ibero, Bacon, {Den Brok}, Dreizler, Kelz, Marino, \&
  Steinmetz}]{Roth2018}
Roth, M.~M., Sandin, C., Kamann, S., {et~al.} 2018, A{\&}A, 618, A3

\bibitem[{Rousseau-Nepton {et~al.}(2019)Rousseau-Nepton, Martin, Robert,
  Drissen, Amram, Prunet, Martin, Moumen, Adamo, Alarie, Barmby, Boselli,
  Bresolin, Bureau, Chemin, Fernandes, Combes, Crowder, {Della Bruna}, {Duarte
  Puertas}, Egusa, Epinat, Ksoll, Girard, {G{\'{o}}mez Llanos}, Gouliermis,
  Grasha, Higgs, Hlavacek-Larrondo, Ho, Iglesias-P{\'{a}}ramo, Joncas, Kam,
  Karera, Kennicutt, Klessen, Lianou, Liu, Liu, {Luiz de Amorim}, Lyman,
  Martel, Mazzilli-Ciraulo, McLeod, Melchior, Millan, Moll{\'{a}}, Momose,
  Morisset, Pan, Pati, Pellerin, Pellegrini, P{\'{e}}rez, Petric, Plana,
  Rahner, {Ruiz Lara}, S{\'{a}}nchez-Menguiano, Spekkens, Stasinska, Takamiya,
  {Vale Asari}, \& V{\'{i}}lchez}]{Rousseau-Nepton2019}
Rousseau-Nepton, L., Martin, R.~P., Robert, C., {et~al.} 2019, MNRAS, 489, 5530

\bibitem[{Santoro {et~al.}(2022)Santoro, Kreckel, Belfiore, Groves, Congiu,
  Thilker, Blanc, Schinnerer, Ho, {Diederik Kruijssen}, Meidt, Klessen,
  Schruba, Querejeta, Pessa, Chevance, Kim, Emsellem, McElroy, Barnes, Bigiel,
  Boquien, Dale, Glover, Grasha, Lee, Leroy, Pan, Rosolowsky, Saito,
  Sanchez-Blazquez, Watkins, \& Williams}]{Santoro2022}
Santoro, F., Kreckel, K., Belfiore, F., {et~al.} 2022, A{\&}A, 658, 1

\bibitem[{{Sarmiento} {et~al.}(2021){Sarmiento}, {Huertas-Company}, {Knapen},
  {S{\'a}nchez}, {Dom{\'\i}nguez S{\'a}nchez}, {Drory}, \&
  {Falc{\'o}n-Barroso}}]{Sarmiento2021}
{Sarmiento}, R., {Huertas-Company}, M., {Knapen}, J.~H., {et~al.} 2021, ApJ,
  921, 177

\bibitem[{{Scheuermann} {et~al.}(2022){Scheuermann}, {Kreckel}, {Anand},
  {Blanc}, {Congiu}, {Santoro}, {Van Dyk}, {Barnes}, {Bigiel}, {Glover},
  {Groves}, {Klessen}, {Kruijssen}, {Rosolowsky}, {Schinnerer}, {Schruba},
  {Watkins}, \& {Williams}}]{Scheuermann2022}
{Scheuermann}, F., {Kreckel}, K., {Anand}, G.~S., {et~al.} 2022, MNRAS, 511,
  6087

\bibitem[{Sutherland {et~al.}(2018)Sutherland, Dopita, Binette, \&
  Groves}]{Sutherland2018}
Sutherland, R., Dopita, M., Binette, L., \& Groves, B. 2018, {MAPPINGS V:
  Astrophysical plasma modeling code}, Astrophysics Source Code Library, record
  ascl:1807.005

\bibitem[{van~der Maaten \& Hinton(2008)}]{VanderMaaten2008}
van~der Maaten, L. \& Hinton, G. 2008, J. Mach. Learn. Res., 9, 2579

\bibitem[{{Walch} {et~al.}(2012){Walch}, {Whitworth}, {Bisbas}, {W{\"u}nsch},
  \& {Hubber}}]{Walch2012}
{Walch}, S.~K., {Whitworth}, A.~P., {Bisbas}, T., {W{\"u}nsch}, R., \&
  {Hubber}, D. 2012, MNRAS, 427, 625

\bibitem[{{Watts} {et~al.}(2024){Watts}, {Cortese}, {Catinella},
  {Fraser-McKelvie}, {Emsellem}, {Coccato}, {van de Sande}, {Brown},
  {Ascasibar}, {Battisti}, {Boselli}, {Davis}, {Groves}, \&
  {Thater}}]{Watts2024}
{Watts}, A.~B., {Cortese}, L., {Catinella}, B., {et~al.} 2024, MNRAS, 530, 1968

\bibitem[{Williams {et~al.}(1994)Williams, de~Geus, \& Blitz}]{Williams1994}
Williams, J., de~Geus, E.~J., \& Blitz, L. 1994, ApJ, 428, 693

\bibitem[{Zhang {et~al.}(2017)Zhang, Yan, Bundy, Bershady, Haffner, Walterbos,
  Maiolino, Tremonti, Thomas, Drory, Jones, Belfiore, S{\'{a}}nchez,
  Diamond-stanic, Bizyaev, Nitschelm, Andrews, Brinkmann, Brownstein, Cheung,
  Li, Law, {Roman Lopes}, Oravetz, Pan, {Storchi Bergmann}, Simmons, Sebastian,
  Diamond-stanic, Bizyaev, Nitschelm, Andrews, Brinkmann, Brownstein, Cheung,
  Li, Law, Lopes, \& Oravetz}]{Zhang2017}
Zhang, K., Yan, R., Bundy, K., {et~al.} 2017, MNRAS, 466, 3217

\end{thebibliography}


\end{document}